\documentclass[twocolumn,preprintnumbers,prl,aps]{revtex4-1}
\usepackage{graphicx,amsmath,bbm,color}
\usepackage[breaklinks=true]{hyperref}
\graphicspath{ {fig/} }

\allowdisplaybreaks[4]

\definecolor{darkblue}{rgb}{0.0,0.0,0.5}
\newcommand{\txblu}{\textcolor{darkblue}}
\newcommand{\Graph}[2]{\vcenter{\hbox{\includegraphics[scale=#1]{#2}}}}
\newcommand{\ff}{\bar{\mathcal{F}}}

\begin{document}

\preprint{MITP/16-097, TCDMATH-16-15}

\title{\boldmath
Quark and gluon form factors to four loop order in QCD: the $N_f^3$ contributions
\unboldmath}

\author{Andreas von Manteuffel\,$^{a,b}$ and Robert M. Schabinger\,$^c$} 

\affiliation{$^a$\,PRISMA Cluster of Excellence, Johannes Gutenberg University, 55099 Mainz, Germany\\
$^b$\,Department of Physics and Astronomy, Michigan State University, East Lansing, Michigan 48824, USA\\
$^c$\,Hamilton Mathematics Institute, Trinity College, Dublin 2, Ireland}

\begin{abstract}
\noindent
We calculate the four-loop massless QCD corrections with three closed quark lines
to quark and gluon form factors.
We apply a novel integration by parts algorithm based on modular arithmetic
and compute all relevant master integrals for arbitrary values of the space-time dimension.
This is the first calculation of a gluon form factor at this perturbative order in QCD.
\end{abstract}

\maketitle

In this Letter, we consider four-loop corrections in massless Quantum Chromodynamics (QCD) to the basic quark and gluon form factors for the photon-quark-antiquark vertex
and the Higgs-gluon-gluon vertex in the infinite top-quark-mass limit, respectively.
They describe the purely virtual QCD corrections to two of the most important Large Hadron Collider processes, the production of a Drell-Yan lepton pair \cite{Drell:1970wh}
and the production of a Higgs boson via gluon fusion \cite{Georgi:1977gs}
in the infinite top-mass limit \cite{Wilczek:1977zn,Shifman:1978zn,Ellis:1979jy,Inami:1982xt}.
Moreover, the poles of the bare loop form factors up to $\mathcal{O}\left(\epsilon^{-2}\right)$ in the $\epsilon$ expansion
\cite{'tHooft:1972fi} allow one to extract the four-loop
cusp anomalous dimensions for the quark and the gluon, respectively.
These quantities enter state-of-the-art resummed predictions
(see \cite{vonManteuffel:2015gxa} and the references therein for more details).

As has long been known, the cusp anomalous dimensions play a central role in the renormalization group evolution of the massless QCD form factors \cite{Magnea:1990zb}
and effectively characterize the leading infrared divergences at $L$-loop order which are not fixed by lower-loop contributions. 
In a recent paper \cite{Henn:2016men}, first results for the four-loop quark form factor and cusp anomalous dimension were presented for a subset of the fermionic terms in the large-$N_c$ limit of the gauge group $SU(N_c)$.
Two other recent four-loop quark cusp anomalous dimension calculations were carried out by computing appropriate matrix elements of soft Wilson lines \cite{Grozin:2015kna} or parton splitting functions \cite{Ruijl:2016pkm}; 
indeed the cusp anomalous dimensions are universal quantities relevant to many aspects of massless QCD and, accordingly, they may be computed in many different ways.
In a related note, we would also like to mention a recent {\it ab initio} calculation of the unintegrated
four-loop form factor in $\mathcal{N}=4$ supersymmetric Yang-Mills theory~\cite{Boels:2012ew,Boels:2015yna}.

Nevertheless, the complete calculation of the four-loop cusp
anomalous dimensions and form factors remains a challenge, both because of the integral
reductions and the calculation of the master integrals.
The gluon form factor is particularly demanding.
The unreduced quark form factor features, at worst, twelve-line integrals with numerator insertions of rank five, but the unreduced gluon
form factor (and, actually, already its $N_f^3$ contributions) has twelve-line integrals with up to rank six numerator insertions. 
In the usual approach to multi-loop form factor calculations, integration by
parts (IBP) identities~\cite{Tkachov:1981wb,Chetyrkin:1981qh} are exploited to reduce
the loop integrals using Laporta's algorithm~\cite{Laporta:2001dd}.
Public computer packages \cite{Smirnov:2014hma,vonManteuffel:2012np,Lee:2012cn} exist, but face significant
technical limitations for the problem of interest.
Laporta's original algorithm involves the solution of a large linear system of equations
with polynomial entries, a task which is well-known to cause both run time and
memory management issues for practical implementations.
New approaches have been discussed~\cite{Gluza:2010ws,Lee:2012cn,Ruijl:2015aca,Ita:2015tya,Larsen:2015ped,Ueda:2016sxw}
to reorganize the IBP identities and thereby allow for a more efficient reduction.

In~\cite{vonManteuffel:2014ixa}, we proposed to improve Laporta's algorithm by
sampling the IBP equations with integer numbers for the variables, employing
modular arithmetic for the reduction step and reconstructing the full rational
solution from sufficiently many such samples.
For the calculation presented in this Letter, we use a new computer program
({\tt Finred}) developed by one of us, which is based on this novel method.
For the reductions of the three-loop form factors, the new program is faster
than {\tt Reduze\;2}~\cite{vonManteuffel:2012np} by two orders of magnitude
on a desktop computer.

In this Letter, we compute the $N_f^3$ contributions to the bare quark and gluon form factor at four-loop order in massless QCD, where $N_f$ is the number of light quark flavors.
For the gluon form factor, this is a new result, while for the quark form factor it is a check of the $N_f^3$ result of~\cite{Henn:2016men}.
After giving a brief general description of our calculation, we present closed-form results for the eight $N_f^3$ master integrals valid for arbitrary values of the parameter of dimensional regularization.
We then provide $\epsilon$-expanded results for the $N_f^3$ form factors through the finite terms and discuss the cross-checks which we carried out to validate our results. 
Finally, we conclude by discussing how the methods that we have developed may allow for the calculation of the still-unknown contributions to the four-loop quark and gluon form factors.

The bare quark and gluon form factors we are interested in, 
$\ff_{\rm bare}^{q}\left(\alpha_s^{\rm bare}, q^2, \mu_\epsilon^2, \epsilon\right)$
and $\ff_{\rm bare}^{g}\left(\alpha_s^{\rm bare}, q^2, \mu_\epsilon^2, \epsilon\right)$,
are given respectively by the interference of the bare three-point functions
for $\gamma^\ast(q)\to q(p_1)\bar{q}(p_2)$ and
the infinite top-mass limit of $h(q) \to g(p_1) g(p_2)$ with the corresponding tree-level expressions.
We employ conventional dimensional regularization and sum over color 
and polarizations of the external particles. The absolute
normalization of our perturbative expansions is precisely that of references \cite{Baikov:2009bg,Gehrmann:2010ue}; we divide our results by the appropriate squared tree-level matrix elements and proceed with the
$\overline{\rm MS}$ renormalization scheme in mind,
\begin{align}
\label{eq:expbareg}
&\ff_{\rm bare}^{q,g}\left(\alpha_s^{\rm bare}, q^2, \mu_\epsilon^2, \epsilon\right) = \\
&\qquad 1 + \sum_{L = 1}^\infty \left(\frac{\alpha_s^{\rm bare}}{4\pi}\right)^L 
\left(\frac{4\pi \mu_\epsilon^2}{-q^2}\right)^{L \epsilon}
e^{-L\epsilon\gamma_E}\ff_L^{q,g}(\epsilon).\nonumber
\end{align}
Here, all partons are treated as massless, $p_1^2=p_2^2=0$, $q^2=(p_1+p_2)^2$ is
the momentum transfer squared, $\alpha_s^{\rm bare}$ the bare strong coupling constant,
$\mu_\epsilon$ the 't Hooft scale, $\gamma_E$ Euler's constant, and $\epsilon=(4-d)/2$
the parameter of dimensional regularization.

The first step of the calculation is to generate all four-loop Feynman diagrams which contribute to the term proportional to $N_f^3$ in the form factors,
$\ff_4^{q}(\epsilon)|_{N_f^3}$ and $\ff_4^{g}(\epsilon)|_{N_f^3}$,
using the program {\tt QGraf} \cite{Nogueira:1991ex} in two different gauges.
One calculation is performed in general $R_\xi$ gauge, where we keep all dependence on $\xi$ and allow for arbitrary reference vectors for the parametrization
of the polarization vectors of the external gluons.
The other calculation uses $\xi = 1$ background field gauge \cite{Abbott:1980hw}, which leads to different interactions and a different number of contributing Feynman diagrams for the gluon form factor.
All diagrams can be matched onto two integral families, one planar and one non-planar, using {\tt Reduze\;2}. 
Once all four-loop diagrams have been appropriately normalized and interfered with the tree-level diagram, the required numerator algebra is carried out in {\tt Form 4} \cite{Kuipers:2012rf,vanRitbergen:1998pn}
for the $R_\xi$ gauge version of the calculation and in {\tt Mathematica} for the background field gauge version of the calculation.

For the IBP reductions of the loop integrals with {\tt Finred}, we used 64~bit
prime numbers both for the finite field modulus and to sample $d$.
The reconstructed reduction identities are tested for correctness by checking the solution obtained for at least five further independent samples.
We find 109 inequivalent planar and non-planar sectors, for which we generate up to $\mathcal{O}\left(10^8\right)$ equations per sector at the outset.
Although ten master integrals occur at intermediate stages of our calculation of the $N_f^3$ terms, two of the master integrals drop out of our final results.
Using the conventions of \cite{vonManteuffel:2015gxa}
with $q^2=-1$ and a normalization of $(\Gamma(1-\epsilon)/(i \pi^{2-\epsilon}))^4$
we find for the remaining eight master integrals
\begin{equation}
\label{A_6_8620}
\Graph{.3}{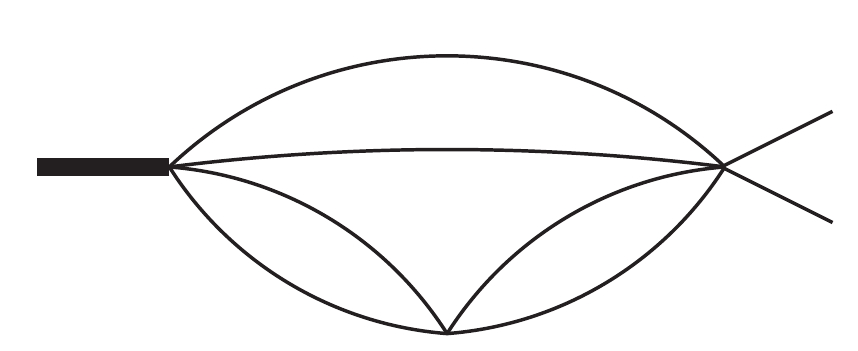} = \frac{\Gamma (2-3 \epsilon ) \Gamma^{10} (1-\epsilon ) \Gamma^2 (\epsilon ) \Gamma (4 \epsilon -2)}{\Gamma (4-5 \epsilon ) \Gamma^2 (2-2 \epsilon ) \Gamma (2 \epsilon )}
\end{equation}
\begin{equation}
\Graph{.3}{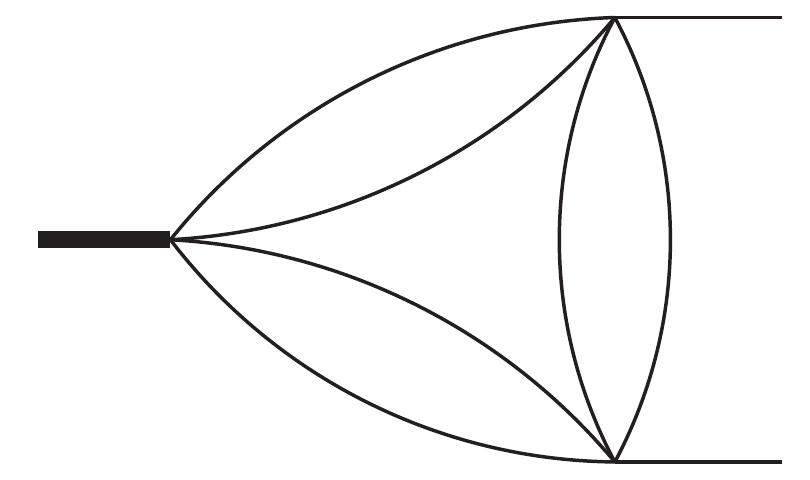} = \frac{\Gamma^2 (2-3 \epsilon ) \Gamma^{10} (1-\epsilon ) \Gamma (\epsilon ) \Gamma (4 \epsilon -2)}{\Gamma (4-5 \epsilon ) \Gamma^3 (2-2 \epsilon )}
\end{equation}
\begin{equation}
\Graph{.3}{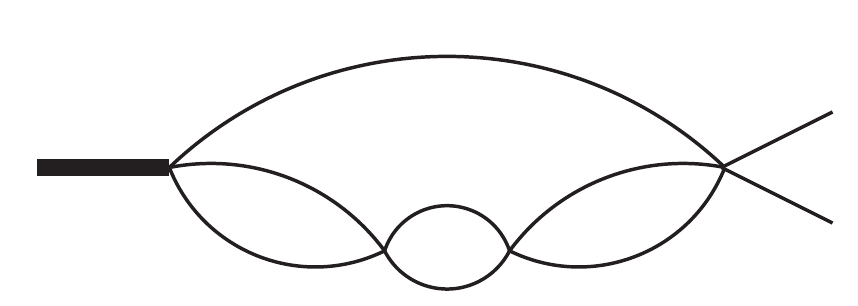} = \frac{-\Gamma (2-4 \epsilon ) \Gamma^{11} (1-\epsilon ) \Gamma^3 (\epsilon ) \Gamma (4 \epsilon -1)}{\Gamma (3-5 \epsilon ) \Gamma^3 (2-2 \epsilon ) \Gamma (3 \epsilon )}
\end{equation}
\begin{align}
&\Graph{.3}{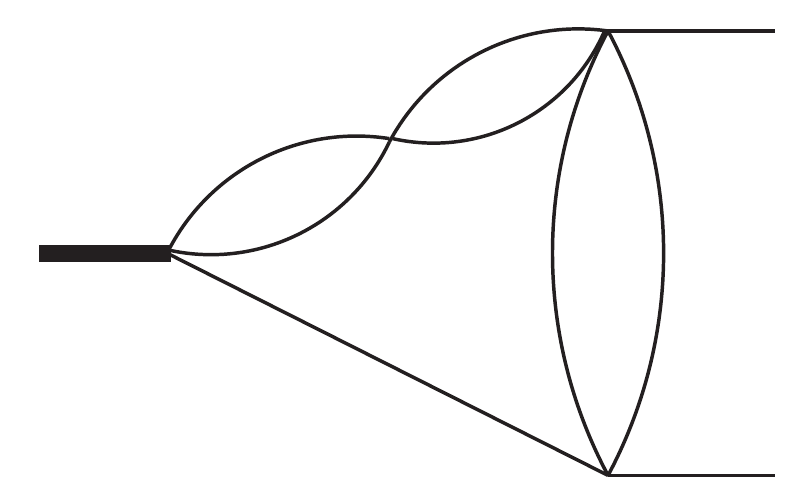} = \nonumber \\
&\qquad \frac{-\Gamma (2-4 \epsilon ) \Gamma (1-2 \epsilon ) \Gamma^{10} (1-\epsilon ) \Gamma^3 (\epsilon ) \Gamma (4 \epsilon -1)}{\Gamma (3-5 \epsilon ) \Gamma^3 (2-2 \epsilon ) \Gamma (2 \epsilon)}
\end{align}
\begin{align}
&\Graph{.3}{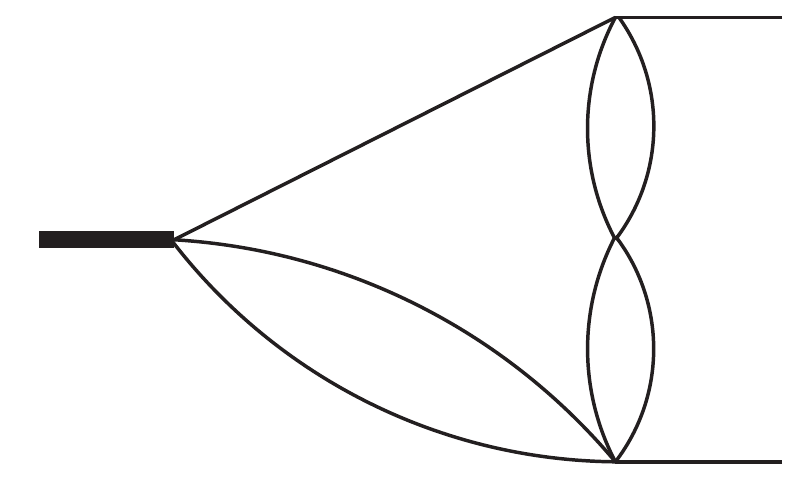} = \nonumber \\
&\qquad \frac{-\Gamma (2-4 \epsilon ) \Gamma (1-3 \epsilon ) \Gamma^{10} (1-\epsilon ) \Gamma^2 (\epsilon ) \Gamma (4 \epsilon -1)}{\Gamma (3-5 \epsilon ) \Gamma^3 (2-2 \epsilon )}
\end{align}
\begin{equation}
\label{A_8_120969}
\Graph{.3}{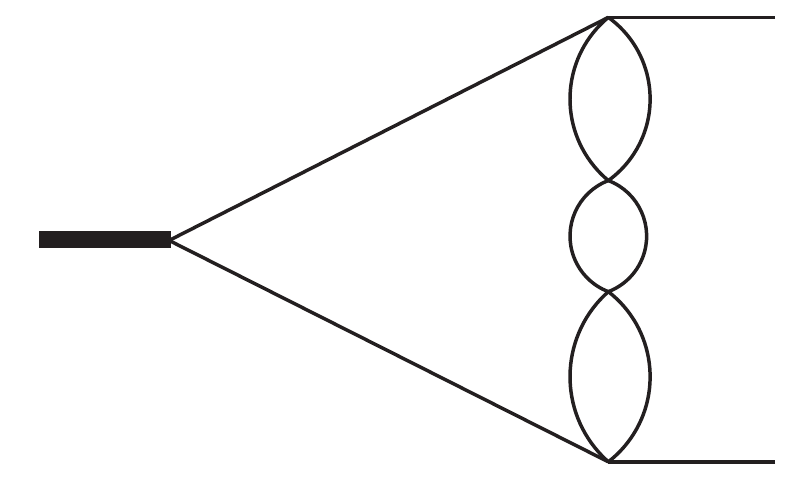} = \frac{\Gamma^2 (1-4 \epsilon ) \Gamma^{10} (1-\epsilon ) \Gamma^3 (\epsilon ) \Gamma (4 \epsilon )}{\Gamma (2-5 \epsilon ) \Gamma^3 (2-2 \epsilon )}
\end{equation}
\begin{widetext}
\begin{align}
\label{D_9_143606}
&\Graph{.35}{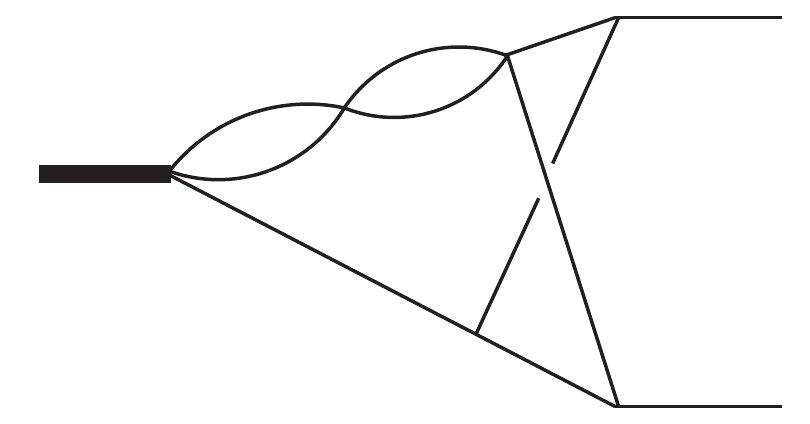} = \frac{2 \Gamma (1-4 \epsilon ) \Gamma (-2 \epsilon ) \Gamma (-\epsilon ) \Gamma^3 (\epsilon ) \Gamma (\epsilon +2) \Gamma (4 \epsilon +1) \Gamma^9 (1-\epsilon ) {}_4 F_3(1,1,\epsilon +2,4\epsilon +1;2,2,2-\epsilon ;1)}
{\Gamma (1-6 \epsilon ) \Gamma^2 (2-2 \epsilon ) \Gamma (2-\epsilon ) \Gamma (2 \epsilon ) \Gamma (\epsilon +1)} \nonumber \\
&+\frac{\Gamma (1-4 \epsilon ) \Gamma (1-2\epsilon ) \Gamma^2 (-\epsilon ) \Gamma^2 (\epsilon ) \Gamma (4 \epsilon ) \Gamma (\epsilon +2) \Gamma^8 (1-\epsilon ) {}_4 F_3(1,1,1-2 \epsilon ,\epsilon +2;2,2,2-\epsilon ;1)}
{\Gamma (1-6\epsilon ) \Gamma^2 (2-2 \epsilon ) \Gamma (2-\epsilon ) \Gamma (2 \epsilon )} \nonumber \\
&+\frac{-\Gamma^2 (1-4 \epsilon ) \Gamma (2-3 \epsilon ) \Gamma^2 (-\epsilon ) \Gamma^2 (\epsilon ) \Gamma (4\epsilon +1) \Gamma^8 (1-\epsilon ) {}_4 F_3(1,1-6 \epsilon ,1-4 \epsilon ,2-3 \epsilon ;2-5 \epsilon ,2-4 \epsilon ,2-4 \epsilon ;1)}
{2 \Gamma (2-5 \epsilon ) \Gamma^2 (2-4 \epsilon )\Gamma^2 (2-2 \epsilon ) \Gamma (2 \epsilon +1)} \nonumber \\
&+\frac{\Gamma (1-4 \epsilon ) \Gamma (-2 \epsilon -1) \Gamma^2 (-\epsilon ) \Gamma^2 (\epsilon ) \Gamma (4 \epsilon ) \Gamma (\epsilon +2)\Gamma (4 \epsilon +1) \Gamma^8 (1-\epsilon ) 
{}_5 F_4(1,1,2-4 \epsilon ,\epsilon +2,4 \epsilon +1;2,2,2-\epsilon ,2 \epsilon +2;1)}{\Gamma (1-6 \epsilon ) \Gamma^2 (2-2 \epsilon ) \Gamma(2-\epsilon ) \Gamma (2 \epsilon ) \Gamma (4 \epsilon -1)} \nonumber \\
&+\frac{-\Gamma^4 (-\epsilon ) \Gamma^2 (\epsilon ) \Gamma (2 \epsilon +1) \Gamma^7 (1-\epsilon ) {}_4 F_3(1-6 \epsilon ,1-\epsilon,-2 \epsilon ,2 \epsilon ;1-3 \epsilon ,1-2 \epsilon ,1-2 \epsilon ;1)}
{4 \Gamma (1-3 \epsilon ) \Gamma^2 (2-2 \epsilon )}
\end{align}
\end{widetext}
\begin{widetext}
\begin{align}
\label{D_8_6620}
\Graph{.3}{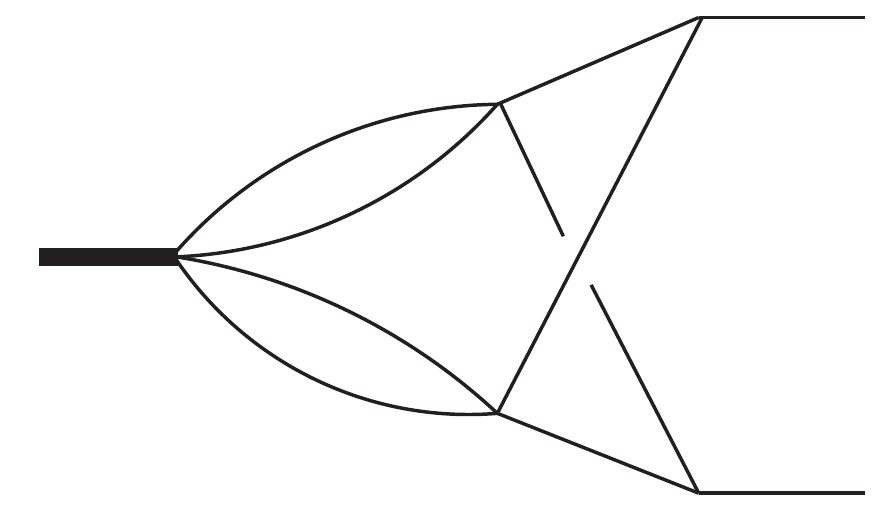} &= \frac{-2 \Gamma^2 (1-3 \epsilon ) \Gamma (-\epsilon ) \Gamma (\epsilon ) \Gamma (4 \epsilon ) \Gamma (\epsilon +2) \Gamma^9 (1-\epsilon ) {}_4 F_3(1,1,4 \epsilon ,\epsilon +2;2,2,2-\epsilon;1)}
{\Gamma (2-6 \epsilon ) \Gamma^2 (2-2 \epsilon ) \Gamma (2-\epsilon ) \Gamma (\epsilon +1)} \nonumber \\
&\quad
+\frac{2 \Gamma^2 (-\epsilon ) \Gamma (\epsilon ) \Gamma^2 (3 \epsilon -1) \Gamma^8 (1-\epsilon) {}_4 F_3(2-6 \epsilon ,1-3 \epsilon ,2-2 \epsilon ,\epsilon ;2-4 \epsilon ,2-3 \epsilon ,2-3 \epsilon ;1)}
{\Gamma (2-4 \epsilon ) \Gamma (2-2 \epsilon ) \Gamma (3 \epsilon )} \nonumber \\
&\quad
+\frac{2\Gamma (2-3 \epsilon ) \Gamma (-3 \epsilon ) \Gamma^2 (-\epsilon ) \Gamma (4 \epsilon ) \Gamma (\epsilon +2) \Gamma^8 (1-\epsilon ) {}_5 F_4(1,1,2-3 \epsilon ,4 \epsilon ,\epsilon+2;2,2,2-\epsilon ,3 \epsilon +1;1)}
{\Gamma (2-6 \epsilon ) \Gamma^2 (2-2 \epsilon ) \Gamma (2-\epsilon )}
\end{align}
\end{widetext}

Eqs. (\ref{A_6_8620})-(\ref{A_8_120969}) are derived by integrating out massless one-loop bubble and one-external-mass one-loop triangle integrals one loop at a time.
The procedure is carried out in practice using an automated {\tt Mathematica} script written by one of us. Eq. (\ref{D_9_143606}) follows from Eq. (13) of \cite{Gehrmann:2006wg} after integrating out two one-loop massless bubble integrals,
whereas Eq. (\ref{D_8_6620}) is a new result of this article. It is straightforward to derive Eq. (\ref{D_8_6620}) by first integrating out two massless one-loop bubble integrals,
using the setup of \cite{Gonsalves:1983nq} for the two-loop crossed form factor integral topology with general indices, and then explicitly carrying out the remaining non-trivial Feynman parameter integrations.
For the $\epsilon$-expansion of the exact expressions~\eqref{A_6_8620}-\eqref{D_8_6620} we employ the software package {\tt HypExp}~\cite{Huber:2005yg}.

We find for the quark form factor
\begin{align}
\label{ffquark}
&\ff_4^{q}(\epsilon)\Big|_{N_f^3} = \txblu{C_F} \bigg[
  \frac{1}{\epsilon^5 }\left(\frac{1}{27}\right)
  +\frac{1}{\epsilon^4}\left(\frac{11}{27}\right)
  +\frac{1}{\epsilon^3}\left(\frac{10}{27}\zeta_2\right. \nonumber\\
& \left. +\frac{254}{81}\right)
  +\frac{1}{\epsilon^2}\left(-\frac{82}{81}\zeta_3 + \frac{110}{27}\zeta_2+\frac{29023}{1458}\right) 
  +\frac{1}{\epsilon}\left( \frac{302}{135} \zeta_2^2 \right. \nonumber\\
& \left. -\frac{902}{81}\zeta_3 + \frac{2540}{81}\zeta_2+ \frac{331889}{2916}
      \right)
  -\frac{2194}{135}\zeta_5 -\frac{820}{81}\zeta_3 \zeta_2\nonumber\\
& +\frac{3322}{135}\zeta_2^2 -\frac{20828}{243}\zeta_3 + \frac{145115}{729} \zeta_2 +\frac{10739263}{17496}
  +\mathcal{O}(\epsilon) \bigg]
\end{align}
and for the gluon form factor
\begin{align}
\label{ffgluon}
&\ff_4^g(\epsilon)\Big|_{N_f^3} = \txblu{C_A}\bigg[
  \frac{1}{\epsilon^5}\left(\frac{1}{27}\right)
 +\frac{1}{\epsilon^4}\left(\frac{5}{27}\right)
 +\frac{1}{\epsilon^3}\left(-\frac{14}{27} \zeta_2\right. \nonumber\\
&\left. -\frac{55}{81}\right)
 +\frac{1}{\epsilon^2}\left(-\frac{586}{81}\zeta_3 -\frac{70}{27}\zeta_2-\frac{24167}{1458}\right)
 + \frac{1}{\epsilon}\left(-\frac{802}{135}\zeta_2^2 \right. \nonumber\\
& \left. -\frac{5450}{81}\zeta_3-\frac{262}{81}\zeta_2  -\frac{465631}{2916}\right)-\frac{14474}{135}\zeta_5+\frac{4556}{81} \zeta_3\zeta_2 \nonumber \\
& -\frac{1418}{27} \zeta_2^2-\frac{99890}{243}\zeta_3+\frac{38489}{729} \zeta_2-\frac{20832641}{17496}+ \mathcal{O}\left(\epsilon\right)\bigg] \nonumber\\
&+\txblu{C_F}\bigg[
\frac{1}{\epsilon^3}\left(-\frac{2}{3}\right)
+\frac{1}{\epsilon^2}\left(\frac{32}{3}\zeta_3-\frac{145}{9}\right)+\frac{1}{\epsilon} \left(\frac{352}{45}\zeta_2^2 \right. \nonumber \\ 
& \left. +\frac{1040}{9}\zeta_3 +\frac{68}{9}\zeta_2 -\frac{10003}{54}\right) +\frac{4288}{27}\zeta_5 -64 \zeta_3 \zeta_2 \nonumber \\
& +\frac{2288}{27}\zeta_2^2 +\frac{24812}{27}\zeta_3 +\frac{3074}{27}\zeta_2 -\frac{508069}{324} + \mathcal{O}\left(\epsilon\right) \bigg],
\end{align}
where $C_A$ and $C_F$ are, respectively, the quadratic Casimir invariants of the adjoint and fundamental representations of the gauge group.

We carried out several cross-checks to validate our results.
First, we obtained identical results in general $R_\xi$ gauge and $\xi = 1$ background field gauge.
In particular, the exact cancellation of all terms depending on $\xi$ or the gluon polarization reference vectors represents a strong check on the reduction identities.
Due to the simplicity of the master integrals relevant to the $N_f^3$ terms,
we could use {\tt FIESTA 4} \cite{Smirnov:2015mct} to check all master integrals through to weight six to part per mille precision or better. 
An important check on the pole terms of $\mathcal{O}\left(\epsilon^{-3}\right)$ and higher
was a comparison to the predictions of the evolution equation. Using Eq. (2.17) of \cite{Moch:2005id}, we find that our higher-order poles have exactly the form required.
Our results for the $\mathcal{O}(\epsilon^{-2})$ poles allow us to extract the
$N_f^3$ contributions to the four-loop quark and gluon cusp anomalous dimensions
\begin{align}
\label{cuspquark}
\Gamma_4^q\Big|_{N_f^3} = \txblu{C_F} \bigg[ \frac{64}{27}\zeta_3 -\frac{32}{81} \bigg]\\
\label{cuspgluon}
\Gamma_4^g\Big|_{N_f^3} = \txblu{C_A} \bigg[ \frac{64}{27}\zeta_3 -\frac{32}{81} \bigg]
\end{align}
which is in agreement with the result of~\cite{Grozin:2015kna}.
Note that the $C_F$ contribution to the gluon form factor arising from non-planar diagrams drops out
for the cusp anomalous dimension as expected from the Wilson loop picture, and Casimir scaling~\cite{Becher:2009qa,Gardi:2009qi},
\begin{equation}
\label{casimirscaling}
\frac{\Gamma_4^q|_{N_f^3}}{\Gamma_4^g|_{N_f^3}} = \frac{\txblu{C_F}}{\txblu{C_A}},
\end{equation}
holds for this class of contributions.
For the quark form factor, we compare our result~\eqref{ffquark} to the $N_f^3$ contribution in \cite{Henn:2016men} and find complete agreement.

Let us conclude by giving a brief outlook for the calculation of the remaining
corrections to the quark and gluon form factors which are still unknown.
We expect the {\tt Finred} program developed to carry out the research described in this Letter to allow for the calculation of the remaining reduction identities.
For the calculation of the master integrals,
a fruitful approach in many cases is to employ a basis of finite integrals~\cite{vonManteuffel:2014qoa}.
In this way, the master integrals become accessible to direct integration methods,
either analytically using {\it e.g.} {\tt HyperInt}~\cite{Panzer:2014caa} or numerically,
see \cite{vonManteuffel:2015gxa} for more details.
A detailed discussion of finite form factor and other Feynman integrals
from a numerical perspective will be given in a forthcoming paper.

{\em Acknowledgments:}
We gratefully acknowledge Hubert Spiesberger, Stephan Weinzierl, and the PRISMA
team for their essential help with acquiring the computational resources needed for this work. We are particularly indebted to Hubert Spiesberger for spearheading these efforts.
Parts of the computations were conducted on the supercomputer Mogon at Johannes Gutenberg University Mainz (\url{www.hpc.uni-mainz.de}), 
and we wish to express our special thanks to the Mogon team for their technical support.
We gratefully acknowledge Thomas Luthe for collaborations on reductions at an early stage of this work,
Lorenzo Tancredi for useful {\tt Form} tips, and Erik Panzer for discussions and related collaborations.
The work of RMS was supported by the European Research Council
through grants 291144 (EFT4LHC) and 647356 (CutLoops).
We are grateful to the Mainz Institute for Theoretical Physics (MITP) for its hospitality and support.
Our figures were generated using {\tt Jaxodraw} \cite{Binosi:2003yf}, based on {\tt AxoDraw} \cite{Vermaseren:1994je}.

\bibliography{ff4lnf3}

\end{document}